\documentclass{sig-alternate-05-2015}
\usepackage{multirow}
\usepackage{mathtools}
\usepackage{stmaryrd}
\usepackage{paralist}
\usepackage{url}
\usepackage{relsize}
\usepackage{multirow}
\usepackage{graphicx}
\usepackage{float}
\usepackage{multicol}
\usepackage{subfigure}
\usepackage{hyperref}
\usepackage{makeidx}  
\usepackage{graphicx}
\usepackage{pgfplots}
\usepackage{multirow}
\usepackage{amsmath}
\usepackage{amssymb}
\usepackage{subfigure}
\usepackage{ifsym}
\usepackage{stmaryrd}
\usepackage{algorithm}
\usepackage{algorithmic}
\usepackage{tikz,times}
\usepackage{xcolor}

\usepackage{lipsum}

\newcommand{\var}[1]{\mathit{vars}(#1)}

\newcommand{\andd}{\mathbin{\mathrm{AND}}}
\newcommand{\OPT}{\mathbin{\mathrm{OPT}}}

\newcommand{\UNION}{\mathbin{\mathrm{UNION}}}

\newcommand{\FILTER}{\mathbin{\mathrm{FILTER}}}
\newcommand{\SELECT}{\mathop{\mathrm{SELECT}}}

\newcommand{\Trace}{\mathit{Trace}}

\newcommand{\voc}{\mathit{adom}}

\newcommand{\axis}{\mathit{axis}}
\newcommand{\self}{\mathit{self}}

\newcommand{\nexts}{\mathit{next}}
\newcommand{\edge}{\mathit{edge}}

\newcommand{\ind}[1]{\mathit{index}(#1)}

\newcommand{\true}{\mathit{true}}
\newcommand{\false}{\mathit{false}}
\newcommand{\error}{\mathit{error}}
\newcommand{\sem}[1]{\llbracket #1 \rrbracket_{G}}
\newcommand{\semm}[2]{\llbracket #1 \rrbracket_{#2}}

\newcommand{\edges}{\mathit{edge}}
\newcommand{\nodes}{\mathit{node}}
\newcommand{\dom}[1]{\mathrm{dom}(#1)}

\newcommand{\adom}[1]{\mathrm{adom}(#1)}
\newcommand{\elt}{\mathit{elt}}

\newcommand{\iri}{\mathit{iri}}

\newcommand{\const}[1]{const(#1)}

\newtheorem{theorem}{Theorem}
\newtheorem{proposition}[theorem]{Proposition}

\newtheorem{lemma}[theorem]{Lemma}
\newtheorem{claim}[theorem]{Claim}

\newtheorem{example}[theorem]{Example}

\usepackage{tikz}
\usetikzlibrary{arrows,positioning,automata,decorations,fit,backgrounds,calc,shapes,snakes}
\usepgflibrary{shapes.geometric} 
\usetikzlibrary{shapes.geometric} 
\usepgflibrary{decorations.pathmorphing} 
\usetikzlibrary{decorations.pathmorphing} 
\usepackage{verbatim}
\usetikzlibrary{calc,backgrounds}
\usetikzlibrary{trees}

\begin{document}

\setcopyright{acmcopyright}

\doi{http://dx.doi.org/10.1145/0000000.0000000}

\isbn{978-1450317412}

\conferenceinfo{AOSD'12}{Hasso-Plattner Institut Potsdam, Germany,March 25--30, 2012}

\acmPrice{\$15.00}

%
\conferenceinfo{XXX}{XXX}
\CopyrightYear{2016} 
\crdata{0-12345-67-8/90/01}  

\title{Path discovery by Querying the federation of Relational Database and RDF Graph}
%
%
%
%
%

\numberofauthors{3} 
%
\author{
%
%
\alignauthor
Xiaowang Zhang$^{1,3}$ \and Jiahui Zhang$^{1,3}$ \and  Muhammad Qasim Yasin$^{2,3}$ \and  {Wenrui Wu$^{1,3}$ \quad\quad Zhiyong Feng$^{2,3}$}\\
       \affaddr{$^1$School of Computer Science and Technology,Tianjin University, Tianjin 300350, P. R. China}\\
       \affaddr{$^2$School of Computer Software,Tianjin University, Tianjin 300350, P. R. China}\\
       \affaddr{$^3$Tianjin Key Laboratory of Cognitive Computing and Application, Tianjin 300350, P.R. China}}

\maketitle
\begin{abstract}
The class of queries for detecting path is an important as those can extract implicit binary relations over the nodes of input graphs. Most of the path querying languages used by the RDF community, like property paths in W3C SPARQL 1.1 and nested regular expressions in nSPARQL are based on the regular expressions. Federated queries allow for combining graph patterns and relational database that enables the evaluations over several heterogeneous data resources within a single query. Federated queries in W3C SPARQL 1.1 currently evaluated over different SPARQL endpoints. In this paper, we present a federated path querying language as an extension of regular path querying language for supporting RDF graph integration with relational database. The federated path querying language is absolutely more expressive than nested regular expressions and negation-free property paths. Its additional expressivity can be used for capturing the conjunction and federation of nested regular path queries. Despite the increase in expressivity, we also show that federated path queries are still enjoy a low computational complexity and can be evaluated efficiently.
\end{abstract}

%
%

%


%
%

%
%

\terms{Theory}


\section{Introduction}\label{sec:intro}
The Resource Description Framework (RDF) \cite{rdfprimer} recommended by World Wide Web Consortium (W3C) is a standard graph-oriented model for interchanging data on the Web \cite{gutierrez_survey}.RDF has implemented in a broad range of applications including the semantic web, social network, bio-informatics, geographical data, etc\cite{abs_book}.
Graph-structured data is typical to access due its navigational nature
\cite{DBLP:conf/icdt/HellingsKBZ13,DBLP:conf/pods/LibkinRV13,DBLP:journals/isci/FletcherGLSBGVW15}.
Navigational path queries on graph databases return binary relations over the nodes of the graph \cite{barcelo_crpq_pods}.
Many existing navigational query languages for graphs are based on binary relational algebra such as XPath
(a standard navigational query language for trees\cite{marxrijke_xpath})
or regular expressions such as RPQ (regular path queries) \cite{DBLP:conf/icdt/Reutter0V15}.

SPARQL \cite{sparql} recommended by W3C has become the standard language for
querying RDF data since 2008 by inheriting classical relational languages
such as SQL. However, SPARQL only provides limited navigational
functionalities for RDF \cite{nsparql,Taski_SPARQL}. Recently, there are
several proposed languages with navigational capabilities for queering RDF graphs
\cite{versa,DBLP:conf/esws/KochutJ07,nsparql,DBLP:journals/ijwis/AlkhateebE14,DBLP:conf/aaai/FiondaPC15,DBLP:conf/semweb/KostylevR0V15}.
Roughly, Versa \cite{versa} is the first language for RDF with navigational
capabilities by using XPath over the XML serialization of RDF graphs.
SPARQLeR proposed by Kochut et al. \cite{DBLP:conf/esws/KochutJ07}  extends
SPARQL by allowing path variables. CPSPARQL proposed by Alkhateeb et al. \cite{DBLP:journals/ijwis/AlkhateebE14} allows
constraints over regular expressions in PSPARQL where variables are allowed in regular expressions. nSPARQL proposed by P\'erez et al. \cite{nsparql} extends SPARQL by allowing nested regular expressions in triple patterns
Indeed, nSPARQL is still expressible in SPARQL if the transitive closure relation is
absent \cite{Taski_SPARQL}. In March 2013, SPARQL 1.1 \cite{sparql1.1}
recommended by W3C allows property paths which strengthen the navigational
capabilities of SPARQL1.0 and
\cite{DBLP:conf/aaai/FiondaPC15,DBLP:conf/semweb/KostylevR0V15} extend property paths by adding some operators such as intersection etc.

\begin{table*}[t]
\centering
\caption{Relational database}\label{tab:cities}
\begin{tabular}{c|c|c|c|c|c|c}
\hline
ID     &Time  &Driver ID &Vehicle ID &Passenger ID &Start Point &End Point\\ \hline
1 &5:30 AM &184 &F583 &D &P3 &P4\\ \hline
2 &6:00 AM &192 &123H &E &P2 &P4\\ \hline
3 &7:59 AM &217 &8E73 &F &P3 &P5\\ \hline
4 &8:15 AM &204 &B398 &A &P3 &P5\\ \hline
5 &8:28 AM &204 &B398 &B &P1 &P4\\ \hline
6 &8:40 AM &204 &B398 &C &P2 &P5\\ \hline
\end{tabular}
\end{table*}

\begin{figure}[h]
\centering
    \tikzstyle{entity} = [rectangle, minimum width=0.4cm, minimum height=0.3cm, text centered, font=\small, draw=black]
    \tikzstyle{final} = [ellipse, minimum width=0.5cm, minimum height=0.25cm, text centered, text = black, font=\small, draw=black]
    \tikzstyle{arrow} = [thick,->,>=stealth, font=\small]

    \begin{tikzpicture}[scale=0.3][node distance=0.3cm]
        \node(entity1)[entity]{P3};
        \node(entity2)[entity,below of=entity1,yshift=-0.5cm]{P1};
        \node(entity3)[entity,below of=entity2,yshift=-0.5cm,xshift=-0.5cm]{P2};
        \node(entity4)[entity,below of=entity2,yshift=-0.5cm,xshift=1cm]{P4};
        \node(entity5)[entity,right of=entity4,xshift=1cm]{P5};

        \draw[arrow](entity1) -- node[anchor=west]{}(entity2);
        \draw[arrow](entity2) -- node[anchor=west]{}(entity3);
        \draw[arrow](entity2) -- node[anchor=west]{}(entity4);
        \draw[arrow](entity3) -- node[anchor=south]{}(entity4);
        \draw[arrow](entity4) -- node[anchor=south]{}(entity5);
    \end{tikzpicture}
\caption{A RDF Graph about the points on the map.}\label{fig:transp}
\end{figure}
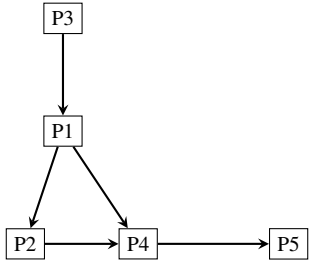

However, those regular expressions-based extensions of SPARQL are still
limited in representing some more expressive navigational queries which are
not expressed in regular expressions.
It let us consider the RDF graph dataset (G) have information about points of longitude and latitude on the map as in Figure \ref{fig:transp}, and a relational database (D) as in Table \ref{tab:cities}.

A record in Table \ref{tab:cities} depicts an order that at some time, a passenger placed to travel from a location to another. In the response of to the order of the passenger, a vehicle is allocated to the driver and asked to fulfil the order of the passenger. But sometime there is no vehicle at the station then the driver with already allocated vehicle near the location of the passenger is supposed to be asked to fulfil the order of the passenger by picking him/her from his location. Which can be possible by querying the federation of relational and RDF (graph data).

Assume that there are three passengers A,B and C. Passenger A has placed an order at 8:15AM that he want to hire a vehicle to travel from point P3 to point P5 as in Table \ref{fig:transp}.At 8:25, the vehicle with Passenger A is near to point P1 and this is recorded .Three minutes later, at 8:28, Passenger B asked for a vehicle and he want to go point 4( P4) from point P1 and this message is stored in Table \ref{tab:cities}.Meanwhile, Passenger C also called for vehicle and he wants to go from point P2 to point P5 at 8:40AM and it recorded in relational database. The system receives the three queries related to same path. By descovery of right path and having information about the vehicle type and time, with one vehicle we can accommodate all of three passenger A, B, and C. as in Figure \ref{fig:transp}, by selecting path ``$\textit{P1} \rightarrow \textit{P2} \rightarrow \textit{P4} \rightarrow \textit{P5}$'' the driver can accommodate the passenger A,B ,C.

Due to the limited space, we omit all proofs in this paper but available in a TR in the link\footnote{\url{http://123.56.79.184/FPQ.pdf}} or arXiv.org.


\section{Preliminaries}\label{sec:preliminaries}
In this section, we briefly recall RDF graphs and the syntax and semantics of nested regular expressions, largely following the excellent exposition \cite{nsparql}.

\subsection{RDF graphs}
An RDF statement is a \emph{subject-predicate-object} structure, called
\emph{RDF triple} which represents resources and the properties of those
resources. For the sake of simplicity similar to \cite{nsparql}, we assume
that RDF data is composed only IRIs\footnote{A standard RDF data is composed
of IRIs, blank nodes, and literals. For the purposes of this paper, the
distinction between IRIs and literals will not be important.}. Formally, let
${U}$ be an infinite set of \emph{IRIs}. A triple $(s, p, o) \in {U} \times
{U} \times {U}$ is called an \emph{RDF triple}.  An \emph{RDF graph} $G$ is
a finite set of RDF triples. We use $\voc(G)$ to denote the \emph{active
domain} of $G$, i.e., the set of all elements from ${U}$ occurring in $G$.

For instance, a RDF graph can be modeled in an RDF graph where each labeled-edge of the form $a \overset{p}{\to} b$ is directly translated into a triple $(a, p, b)$.

Let $G$ be an RDF graph. A \emph{path} $\pi = (c_1c_2 \ldots c_m)$ in $G$ is
a non-empty finite sequence of constants from $G$, where, for every $i \in
\{1, 2,\ldots, m-1\}$, $c_i$ and $c_{i+1}$ exactly occur in the same triple of $G$
(i.e., $(c_{i}, c, c_{i+1})$, $(c_{i}, c_{i+1}, c)$, and $(c, c_{i},
c_{i+1})$ etc.). Note that the precedence between $c_i$ and $c_{i+1}$ in a
path is independent of the positions of $c_i, c_{i+1}$ in a triple.

To capture all binary relations on triples, three different \emph{navigation axes}, namely, $\nexts$, $\edges$, and $\nodes$, and their inverses, i.e., $\nexts^{-1}$,
$\edges^{-1}$, and $\nodes^{-1}$, are introduced to move through an RDF triple $(s, p, o)$.

Let $\Sigma=\{\axis, {\axis::c} \mid c \in {U}\}$ where $\axis \in \{\self$,
$\nexts$, $\edges$, $\nodes$, $\nexts^{-1}$, $\edges^{-1}$,
$\nodes^{-1}\}$. Let $G$ be an RDF graph. We use $\Sigma(G)$ to denote the
set of all symbols $\{\axis, {\axis::c} \mid c \in \voc(G)\}$ occurring in
$G$.

Let $\pi = (c_1\ldots c_m)$ a path in $G$. A
\emph{trace} of path $\pi$ is a string over $\Sigma(G)$ written by
$\mathcal{T}(\pi) = l_1\ldots l_{m-1}$ where, for all $ i \in \{1, \ldots,
m-1\}$, $(c_ic_{i+1})$ is labeled by $l_i$ in the following manner: let
$\axis \in \{\nexts, \edges, \nodes\}$,
\begin{compactitem}
\item $l_i = \self$ if $c_{i} = c_{i+1}$;
\item $l_i = \self::c_{i}$ if $c_{i} = c_{i+1}$;
\item $l_i = \nexts::c$ if $(c_{i}, c, c_{i+1}) \in G$;
\item $l_i = \edges::c$ if $(c_{i}, c_{i+1}, c) \in G$;
\item $l_i = \nodes::c$ if $(c, c_{i}, c_{i+1}) \in G$;
\item $l_i = \nexts$ if $(c_{i}, c,
    c_{i+1}) \in G$ for some $c \in \voc(G)$;
\item $l_i = \edges$ if $(c_{i},
    c_{i+1}, c) \in G$ for some $c \in \voc(G)$;
\item $l_i = \nodes$ if $(c, c_{i},
    c_{i+1}) \in G$ for some $c \in \voc(G)$;
\item $l_i = \axis^{-1}$ if $(c_{i+1}c_i)$ is labeled by $\axis$;
\item $l_i = \axis^{-1}::c$ if $(c_{i+1}c_i)$ is labeled by $\axis::c$.
\end{compactitem}

We use $\Trace(\pi)$ to denote the set of all traces of $\pi$.

Note that it is possible that a path has multiple traces since any two nodes possibly
occur in the multiple triples. For example, an {RDF} graph $G = \{(a, b,
c), (a, c, b)\}$ and given a path $\pi = (abc)$, both
$(\edges::c)(\nodes::a)$ and $(\nexts::c)(\nodes^{-1}::a)$ are traces of
$\pi$.


\subsection{Nested regular expressions}
Nested regular expressions (\emph{nre}) are defined by the following formal
syntax:
\begin{equation*}\label{def:nre}
e:= \axis \mid \axis::c\, (c \in U) \mid \axis::[e] \mid e /e \mid e | e  \mid  e^{\ast}.
\end{equation*}
Here the \emph{nesting nre-expression} is of the form $\axis::[e]$.

Given an RDF graph $G$, the evaluation of $e$ on $G$, denoted by $\semm
{e}{G}$, is a binary relation inductively defined as follows:
\begin{equation*}
\begin{array}{rl}
 \sem {\self} &  = \{(c, c) \mid c \in \voc(G)\};  \\
 \sem {\self::c} & = \sem {\self} \cap \{(c, c)\};  \\
 \sem {\nexts} &  = \{(a, b) \mid \exists\, c,\, (a, c, b) \in G\};  \\
 \sem {\nexts::c} &  =  \{(a, b) \mid (a, c, b) \in G\};  \\
 \sem {\edges} &  = \{(a, b) \mid \exists \, c,\, (s, t, c) \in G\};  \\
 \sem {\edges::c} &  = \{(a, b) \mid (a, b, c) \in G\};  \\
 \sem {\nodes} &  =  \{(a, b) \mid \exists \, c,\, (c, a, b) \in G\};  \\
 \sem {\nodes::c} &  =  \{(a, b) \mid (c, a, b) \in G\};  \\
\sem {\axis^{-1}} & =  \{(a, b) \mid (b, a) \in \sem {\axis}\};\\
\sem {\axis^{-1}::c} & = \{(a, b) \mid (b, a) \in \sem {\axis::c}\};\\
\sem {e_{1} | e_{2}} & =   \sem {e_{1}} \cup \sem {e_{2}};\\
\sem {e_{1}/ e_{2}} & =  \{(a, b) \mid \exists\, c,\, (a, c) \in \sem {e_{1}} \wedge (c, b) \in \sem {e_{2}}\};\\
\sem {e^{\ast}} & =  \sem {\self} \cup \sem {e} \cup \sem {e/e} \cup \sem {e/e/e} \cup \ldots \\
\sem {\self::[e]} &  =  \{(a, a) \mid \exists\, c,\, (a, c) \in \sem {e}\};  \\
\sem {\axis::[e]} &  =  \{(a, b) \mid \exists\, c, d,\, (a, b) \in \sem{\axis::c} \\
& \quad\quad\quad\quad\quad\quad\quad\quad\quad\quad\quad\quad \wedge (c, d) \in \sem {e}\}.
\end{array}
\end{equation*}


\paragraph{Query evaluation}
Let $V$ be a set of variables, disjoint with $U$. It is a SPARQL convention to prefix each variable with a question mark ``?''.

An \emph{nre-triple pattern} is of the form $(u, e, v)$ where $u, v \in U\cup V$ and $e$ is an nre. Given an RDF graph $G$, the semantics of $(u, e, v)$ is defined as follows:
\begin{equation*}
\sem {(u, e, v)} = \{\mu \colon \{u, v\}\cap V \to U \mid (\mu(u),\mu(v)) \in \semm{e}{G}\}.
\end{equation*}
Here, for any mapping $\mu$ and any constant $c \in {U}$, we
agree that $\mu(c)$ equals $c$ itself.

A \emph{nested regular path query} (NRPQ) $\mathbf{q}(u, v)$ is of the form $(u', e, v')$ where
\begin{compactitem}
\item $\mathbf{q}$ is the name of NRPQ;
\item $\{u, v\}\cap V \subseteq \{u', v'\} \cap V$;
\item $(u', e, v')$ is an nre-triple pattern.
\end{compactitem}

Given an RDF graph $G$, an NRPQ $\mathbf{q}(u, v)$, and a mapping $\mu$, the query evaluation problem is deciding whether $\mu$ is in $\semm {\mathbf{q}(u, v)}{G}$. The complexity of query evaluation problem is in time $O(|G|\cdot|\mathbf{q}|)$ \cite{nsparql}.

\section{Conjunctive NRPQ}\label{sec:CNRPQ}
In this section, we introduce an extension of nested regular path queries named \emph{conjunctive nested regular path queries} (CNRPQ).

\subsection{Syntax and semanctics}
In syntax, the conjunctive NRPQ extends NRPQ in a natural way.

Formally, an CNRPQ is of the form $\mathbf{q}(u, v)$ defined as follows:
\begin{equation}\label{equ:CNRE}
\mathbf{q}(u, v) := \bigwedge^{n}_{i=1}\, (u_i, e_i, v_i);\footnote{In this paper, we simply write a conjunctive query as a Datalog rule \cite{abs_book}.}
\end{equation}
where
\begin{compactitem}
    \item $\mathbf{q}$ is the name of CNRPQ;
	\item $\{u, v\}\cap V \subseteq \{u_1, \ldots, u_n, v_1, \ldots, v_n\} \cap V$;
	\item each $(u_i, e_i, v_i)$ for $i \in \{1, \ldots, n\}$ is an nre-triple pattern.
\end{compactitem}

Note that the first item ensures that each CNRPQ is \emph{bounded}, that is, all variables in $u, v$ also occurs in some subqueries of the CNRPQ. And the second item states that all nre-triple patterns of CNRPQ are NRPQ. By default, if both $u$ and $v$ are constants, i.e., $u, v \in U$, then CNRPQ returns the empty mapping $\mu_{\emptyset}$, that is, a mapping with the empty domain. In this sense, CNRPQ is taken as a boolean query, where either ``$\true$'' or ``$\false$'' is returned.

For instance, let $Q(?x, ?y) = (?x, \nexts::\emph{father}, ?z) \wedge (?z, \nexts::\emph{father}, ?y)$ be a CNRPQ. Clearly, $Q$ represents the ``\emph{grandfather}'' relationship.

Semantically, let $\mathbf{q}(u, v)$ be a CNRPQ of the form (\ref{equ:CNRE}) and $G$ be an RDF graph, $\semm {\mathbf{q}(u, v)}{G}$ returns a set of mappings with the domain $\{u, v\} \cap V$ defined as follows:
\begin{multline*}
\{\mu|_{\{u, v\}\cap V} \mid \mu = \mu_1 \cup \mu_2 \cup \ldots \cup \mu_n \text{ and } \\
\forall\, i \in \{1, \ldots, n\},\, \mu_i \in \semm {(u_i, e_i, v_i)}{G}\}.
\end{multline*}

Intuitively, each mapping $\mu$ of $\mathbf{q}(u, v)$ on $G$ is the restriction of $\mu_1 \cup \mu_2 \cup \ldots \cup \mu_n$ where each $\mu_i$ on $\{u, v\}\cap V$ is a mapping of
a subquery $\mathbf{q}_i(u_i, v_i)=(u_i, e_i, v_i)$ for $i =1,2,\ldots, n$.


\begin{example}\label{exam:cnre-nre}
Let $G = \{(a, p, b), (b, q, c), (a, r, c)\}$ (shown in Fig. \ref{fig:cnre-nre-G}) and $H = \{(a, p, b), (b, q, c), (a, r, d)\}$ (shown in Fig. \ref{fig:cnre-nre-H}) be two RDF graphs.
\begin{figure}[h]
\centering
\begin{minipage}[t]{0.5\linewidth}
\scalebox{0.6}{
\begin{tikzpicture}[->,>=stealth',shorten >=1pt,auto,node distance=2cm,
			thick,main node/.style={circle,fill=white!20,draw,font=\sffamily\Large\bfseries}]
			
			\node[main node] (1) {a};
			\node[main node] (2) [right of=1] {b};
			\node[main node] (3) [right of=2] {c};

			\path[every node/.style={font=\sffamily\large}]
			(1) edge node  {$p$} (2)
			(2) edge node {$q$} (3)
			(1) edge [bend right] node [below] {$r$} (3);
			\end{tikzpicture}
}
\caption{G}\label{fig:cnre-nre-G}
\end{minipage}%
\begin{minipage}[t]{0.5\linewidth}
\scalebox{0.6}{
\begin{tikzpicture}[->,>=stealth',shorten >=1pt,auto,node distance=2cm,
			thick,main node/.style={circle,fill=white!20,draw,font=\sffamily\Large\bfseries}]
			
			\node[main node] (1) {a};
			\node[main node] (2) [ right of=1] {b};
			\node[main node] (3) [ right of=2] {c};
			\node[main node] (4) [ below of=2] {d};
			
			\path[every node/.style={font=\sffamily\large}]
			(1) edge node  {$p$} (2)
			(2) edge node  {$q$} (3)
			(1) edge node [below] {$r$} (4);
\end{tikzpicture}
}
\caption{H}\label{fig:cnre-nre-H}
 \end{minipage}%
\end{figure}
	
Consider a CNRPQ $\mathbf{q}(?x, ?y) = (?x, (\nexts::p)/(\nexts::q), ?y) \wedge (?x, (\nexts::r), ?y)$. We have $\semm {\mathbf{q}(?x, ?y)}{G} = \{(?x = a, ?y = c)\}$ while $\semm {Q(?x, ?y)}{H} = \emptyset$.
\end{example}

In other words, Example \ref{exam:cnre-nre} shows that the query $Q$ can distinguish graph $G$ from $H$. However, we find that there exists no any NRPQ to distinguish graph $G$ from $H$ in the following subsection.

\subsection{CNRPQ is not expressible in NRPQ}
In this subsection, we theoretically show that CNRPQ has more expressive power than NRPQ. Firstly, we define the notion of \emph{expressiveness} between two query languages.

Let $L_1$ and $L_2$ be two query languages on RDF graphs. We say $L_1$ is \emph{expressible} in $L_2$ if for any query $\mathbf{q}$, there exists some query $\mathbf{q}'$ for any RDF graph $G$ such that $\semm {\mathbf{q}}{G} = \semm {\mathbf{q}'}{G}$.

Secondly, we introduce an extension of nre nre($\cap$) by adding the intersection operator $\cap$ in nre and then we will show that nre($\cap$) can express the intersection of nre-expressions. Finally, we show that the intersection of nre-expressions is not expressed by any nre-expression.

Let $e_1$ and $e_2$ be two nre-expressions. We use $e_1 \cap e_2$ to denote the \emph{intersection} of $e_1$ and $e_2$.  The evaluation of $e_1 \cap e_2$ is defined as follows: let $G$ be an RDF graph,
\begin{equation}
\semm {e_1\cap e_2}{G}:= \semm {e_1}{G} \cap \semm {e_2}{G}.
\end{equation}
Analogously, we could define NRPQ$^{\cap}$ corresponding to nre($\cap$).

Next, we will show that nre($\cap$) is not expressible in nre.

An RDF graph $G$ is called \emph{p}-RDF graph if all predicates in all triples of $G$ are \emph{p} and neither subject nor object is \emph{p}.
Let $G$ be a \emph{p}-RDF graph. An \emph{induced graph} of $G$ written by $\ind{G}$ is a node-labeled undirected graph obtained from $G$ in the following way:

let $\ind{G} = (V, E, \lambda)$,
\begin{compactitem}
	\item $V(G) = V_1 \cup V_2$ where $V_1 = \{v_a, w_b\mid (a, p, b) \in G\}$ and $V_2= \{u_{ab}\mid (a, p, b) \in G\}$;
	\item $E(G) = \{(v_a, u_{ab}), (u_{ab}, w_b) \mid (a, p, b) \in G\}$;
	\item $\lambda(v_a) = a$, $\lambda(u_{ab}) = p$, and $\lambda(w_b) = b$;
	\item $\lambda(v_1) = \lambda(v_2)$ implies $v_1 = v_2$ for $v_1, v_2 \in V_1$.
\end{compactitem}

Clearly, for every \emph{p}-RDF graph, its all induced graphs are isomorphic.

A \emph{p}-RDF graph $G$ is called \emph{strongly acyclic} if $\ind{G}$ is acyclic.

For instance, the \emph{p}-RDF graph $\{(a, p, b)\}$ is strongly acyclic.

We use nre$^{\mathrm{cf}}$ to denote the \emph{constant-free nre}, that is, $\axis::c$ is free.
\begin{lemma}\label{lem:nre-plus}
For any nre$^{\mathrm{cf}}$ expression $e$, if $(a, b) \in \semm {e}{G}$ for some \emph{p}-RDF graph $G$ and some pair $(a, b)$ with $a, b \in \const{G}$ then there exists some strongly acyclic \emph{p}-RDF graph $H$ such that $(a, b) \in \semm {e}{H}$.
\end{lemma}

The following property shows that the intersection of nre-expressions cannot be expressed by any nre-expression.
\begin{proposition}\label{prop:intersection-nre}
nre$(\cap)$ is not expressible in nre.
\end{proposition}
%
%

By Proposition \ref{prop:intersection-nre}, we can conclude an important result.
\begin{theorem}\label{thm:CNRPQ-NRPQ}
CNRPQ is not expressible in NRPQ.
\end{theorem}

\section{Federated path queries}\label{sec:FPQ}
In this section, we introduce two extensions of conjunctive nested regular path queries named \emph{federated conjunctive nested regular path queries} (FCNRPQ) and \emph{union of federated conjunctive nested regular path queries} (UFCNRPQ) for heterogeneous databases with RDF graphs and relational databases.
\subsection{FCNRPQ}
Let $\mathcal{R}$ be a set of relation names. An FCNRPQ is of the form $\mathbf{q}(u, v)$ defined as follows:
\begin{equation}\label{equ:FCNRE}
\mathbf{q}(u, v) := \varphi \wedge \bigwedge^{n}_{i=1}\, (u_i, e_i, v_i);
\end{equation}
where
\begin{compactitem}
    \item $\mathbf{q}$ is the name of FCNRPQ;
	\item $\{u, v\}\cap V \subseteq (\{u_1, \ldots, u_n, v_1, \ldots, v_n\} \cup \var{\varphi}) \cap V$;
	\item each $(u_i, e_i, v_i)$ for $i \in \{1, \ldots, n\}$ is an nre-triple pattern;
    \item $\varphi$ is a conjunction combination of literals $R(w_1, \ldots, w_m)$ defined as follows:
    \begin{equation*}
    \varphi:= R(w_1, \ldots, w_m) \mid \varphi_1 \wedge \varphi_2.
    \end{equation*}
    Here
    \begin{compactitem}
    \item $R$ is a relation name;
    \item $\{w_1, \ldots, w_m\}  \subseteq  V \cup U$;
    \item $\var{\varphi}$ is the collection of all variables occurring in $\varphi$.
    \end{compactitem}
\end{compactitem}

Intuitively speaking, FCNRPQ is an extension of CNRPQ by introducing the conjunctive queries on relations. By default, we allow $\mathbf{q}(u,v) = \varphi$, that is, CNRPQ is absent. In this case, FCNRPQ is taken as conjunctive first-order logic queries \cite{abs_book}. Besides, CNRPQ is also taken as a fragment of FCNRPQ.

Semantically, let $\mathbf{q}(u, v)$ be an FCNRPQ of the form (\ref{equ:FCNRE}) and $\mathbb{D}= (G, \mathcal{D})$ be a heterogeneous database where $G$ is an RDF graph and $\mathcal{D}$ is a set of relations, $\semm {\mathbf{q}(u, v)}{G}$ returns a set of mappings defined as follows:
\begin{multline*}
\{\mu|_{\{u, v\}\cap V} \mid \mu = \mu_1 \cup \mu_2 \cup \ldots \cup \mu_n \text{ and }
\mu|_{\var{\varphi}} \in \semm {\varphi}{\mathcal{D}}\\ \text{ and }
\forall\, i \in \{1, \ldots, n\},\, \mu_i \in \semm {(u_i, e_i, v_i)}{G}\}.
\end{multline*}
Here $\semm {\varphi}{\mathcal{D}}$ is defined in the following inductive way:
\begin{compactitem}
\item Basically, let $R^{\mathcal{D}}$ be a relation of $\mathcal{D}$ mapped to $R$,
\begin{multline*}
\semm {R(w_1, \ldots, w_m)}{\mathcal{D}} = \{\mu \mid \dom{\mu} = \{w_1, \ldots, w_m\} \cap V \\ \text{ and } (\mu(w_1), \ldots, \mu(w_m)) \in R^{\mathcal{D}}\};
\end{multline*}
\item Inductively, $\semm {\varphi_1 \wedge \varphi_2}{\mathcal{D}} = \semm {\varphi_1}{\mathcal{D}} \Join \semm {\varphi_2}{\mathcal{D}}$, where $ \Omega_1 \Join \Omega = \{\mu_1 \cup \mu_2 \mid \mu_1 \in \Omega_1,\ \mu_2 \in \Omega_2, \ \mu_1 \sim \mu_2\} $ for any two sets of mappings $\Omega_1$ and $\Omega_2$. Here, two mappings $\mu_1$ and $\mu_2$ are \emph{compatible} \cite{perez_sparql_tods}, written by $\mu_1 \sim \mu_2$, if for every variable $?x \in \dom {\mu_1} \cap \dom {\mu_2}$, $\mu_1(?x) = \mu_2(?x)$.
\end{compactitem}


In the following, we will show that FCNRPQ has more expressive power than CNRPQ.

To do so, we introduce the following lemma.
\begin{lemma}\label{lem:ecnrpq-cnrpq}
For any CNRPQ $\mathbf{q}(?x, ?y)$, for any RDF graph $G$ with $G = \{(a, a, a)\}$, if  $a$ does not occur in $\mathbf{q}(?x, ?y)$ then\\ $\semm {\mathbf{q}(?x, ?y)}{G}\neq \emptyset$.
\end{lemma}

By Lemma \ref{lem:ecnrpq-cnrpq}, we can conclude an important result.
\begin{theorem}\label{thm:ecnrpq-cnrpq}
FCNRPQ is not expressible in CNRPQ.
\end{theorem}
%

\subsection{Union of FCNRPQ}
A UFCNRPQ is of the form $Q(u, v)$ defined as follows:
\begin{equation}\label{equ:UFCNRE}
\mathbf{q}(u, v) := \bigvee^{n}_{i=1}\, \mathbf{q}_i(u, v);
\end{equation}
where
\begin{compactitem}
    \item $\mathbf{q}$ is the name of UFCNRPQ;
	\item each $\mathbf{q}_i(u, v)$ is an ECNPRQ for $i\in \{1, 2, \ldots, n\}$.
\end{compactitem}

Semantically, let $\mathbf{q}(u, v)$ be a UFCNRPQ of the form (\ref{equ:UFCNRE}) and $\mathbb{D}= (G, \mathcal{D})$ be a heterogeneous database where $G$ is an RDF graph and $\mathcal{D}$ is a set of relations, $\semm {\mathbf{q}(u, v)}{G}$ returns a set of mappings defined as follows:
\vspace{-2mm}
\begin{equation*}
\semm {\mathbf{q}(u, v)}{G} = \bigcup^{n}_{i=1} \semm {\mathbf{q}_i(u, v)}{\mathbb{D}}.
\end{equation*}


In the following, we will show that UFCNRPQ has more expressive power than FCNRPQ.

%
%

\begin{lemma}\label{lem:fcnrpqp-single}
For any FCNRPQ $\mathbf{q}(?x, ?y)$, for any heterogeneous database $\mathbb{D} = (G, \emptyset)$, if $G$ is a singleton then $\semm{\mathbf{q}(?x, ?y)}{\mathbb{D}}$ contains at most one mapping.
\end{lemma}

\begin{theorem}\label{thm:uecnrpq-ecnrpq}
UFCNRPQ is not expressible in FCNRPQ.
\end{theorem}
%

\section{Expressiveness of FPQ}\label{sec:expressivity}
In previous sections, our proposed path queries NRPQ, CNRPQ, FCNRPQ, and UFCNRPQ are called \emph{federated path queries} (FPQ). In this section, we investigate the expressiveness of FPQ.
\subsection{Expressiveness of variants of RPQ}
To discuss subtly, we introduce some interesting fragments of nre as follows \cite{cfSPARQL}:
\begin{compactitem}
\item nre$_{0}$: \emph{basic nre}, i.e., nre only consisting of ``$\axis$'',
    ``$/$'', and ``$\ast$'';
\item nre$_{0}(|)$: basic nre by adding the operator ``$\mid$'';
\item nre$_{0}(\mathrm{\mathsmaller N})$ to basic nre by adding nesting nre
    $\axis::[e]$.
\end{compactitem}

According to the three fragments of nre, namely, nre$_{0}$, nre$_{0}(|)$, nre$_{0}(\mathrm{\mathsmaller N})$, we can introduce the following three fragments of NRPQ as follows:
\begin{compactitem}
\item RPQ: an NRPQ with nre$_{0}$-expressions;
\item RPQ($|$): an NRPQ with nre$_{0}(|)$-expressions;
\item RPQ($\mathrm{\mathsmaller N}$): an NRPQ with nre$_{0}(\mathrm{\mathsmaller N})$-expressions.
\end{compactitem}
In this sense, NRPQ can be denoted as RPQ($|$,$\mathrm{\mathsmaller N}$).

Analogously, all FPQs can be denoted as RPQ($\mathcal{X}$) where $\mathcal{X}$ is a set of operators such as $|$, $\mathrm{\mathsmaller N}$, $\wedge$, $\mathrm{\mathsmaller R}$, and $\vee$ as follows:
\begin{compactitem}
\item $\wedge$: the conjunctive operator;
\item $\mathrm{\mathsmaller R}$: the federated operator;
\item $\vee$: the union operator.
\end{compactitem}

Thus we can denote CNRPQ, FCNRPQ, and UFCNRPQ as follows:
\begin{compactitem}
\item CNRPQ: RPQ($|$,$\mathrm{\mathsmaller N}$,$\wedge$);
\item FCNRPQ: RPQ($|$,$\mathrm{\mathsmaller N}$,$\wedge$,$\mathrm{\mathsmaller R}$);
\item UFCNRPQ: RPQ($|$,$\mathrm{\mathsmaller N}$,$\wedge$,$\mathrm{\mathsmaller R}$,$\vee$).
\end{compactitem}

By the proofs of Theorem \ref{thm:CNRPQ-NRPQ}, Theorem \ref{thm:ecnrpq-cnrpq}, and Theorem \ref{thm:uecnrpq-ecnrpq}, we can show that the conjunctive operator ($\wedge$),  the federated operator ($\mathrm{\mathsmaller R}$), and the union operator ($\vee$) are primitive. So we can conclude that each fragment with the operator is not expressible in any fragment without the operator \cite{primitive}. That is, RPQ($\mathcal{X}\cup \{\circ\}$) is not expressible in RPQ($\mathcal{X} - \{\circ\}$) where $\circ$ is the placeholder of ``$\wedge$'', ``$\vee$'', or ``$\mathrm{\mathsmaller R}$''.

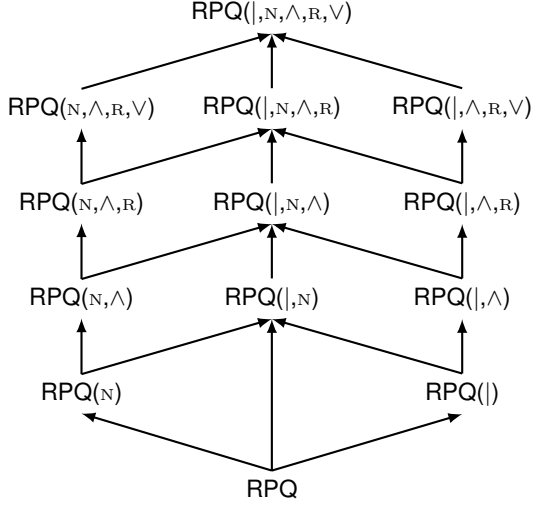
\begin{figure}[h]
\centering
\begin{tikzpicture}[>=latex]
  %
  %
  \tikzstyle{state} = [minimum height=1.5em, minimum width=1.5em, node distance=4em, font={\sffamily}]
  \tikzstyle{stateEdgePortion} = [black,thick];
  \tikzstyle{stateEdge} = [stateEdgePortion, ->];
  \tikzstyle{stateEdgeD} = [stateEdgePortion,dashed,->];
  \tikzstyle{stateEdgeUD} = [stateEdgePortion,dashed];
  \tikzstyle{edgeLabel} = [pos=0.5, text centered, font={\sffamily\small}];

  %
  %
  \node[state, name=p1] {RPQ($|$,$\mathrm{\mathsmaller N}$,$\wedge$,$\mathrm{\mathsmaller R}$,$\vee$)};
  \node[state, name=p3, below of=p1, xshift = -8em] {RPQ($\mathrm{\mathsmaller N}$,$\wedge$,$\mathrm{\mathsmaller R}$,$\vee$)};
  \node[state, name=p4, below of=p1] {RPQ($|$,$\mathrm{\mathsmaller N}$,$\wedge$,$\mathrm{\mathsmaller R}$)};
  \node[state, name=p5, below of=p1, xshift = 8em] {RPQ($|$,$\wedge$,$\mathrm{\mathsmaller R}$,$\vee$)};
  \node[state, name=p6, below of=p3] {RPQ($\mathrm{\mathsmaller N}$,$\wedge$,$\mathrm{\mathsmaller R}$)};
  \node[state, name=p7, below of=p4] {RPQ($|$,$\mathrm{\mathsmaller N}$,$\wedge$)};
  \node[state, name=p8, below of=p5] {RPQ($|$,$\wedge$,$\mathrm{\mathsmaller R}$)};
  \node[state, name=p9, below of=p6] {RPQ($\mathrm{\mathsmaller N}$,$\wedge$)};
  \node[state, name=p10, below of=p7] {RPQ($|$,$\mathrm{\mathsmaller N}$)};
  \node[state, name=p11, below of=p8] {RPQ($|$,$\wedge$)};
  \node[state, name=p12, below of=p9] {RPQ($\mathrm{\mathsmaller N}$)};
  \node[state, name=p13, below of=p10] {};
  \node[state, name=p14, below of=p11] {RPQ($|$)};
  \node[state, name=p15, below of=p13] {RPQ};

  \draw ($(p3.north)$) edge[stateEdge] ($(p1.south)$);
  \draw ($(p4.north)$) edge[stateEdge] ($(p1.south)$);
  \draw ($(p5.north)$) edge[stateEdge] ($(p1.south)$);
  \draw ($(p6.north)$) edge[stateEdge] ($(p3.south)$);
  \draw ($(p6.north)$) edge[stateEdge] ($(p4.south)$);
  \draw ($(p7.north)$) edge[stateEdge] ($(p4.south)$);
  \draw ($(p8.north)$) edge[stateEdge] ($(p4.south)$);
  \draw ($(p8.north)$) edge[stateEdge] ($(p5.south)$);
  \draw ($(p9.north)$) edge[stateEdge] ($(p6.south)$);
  \draw ($(p9.north)$) edge[stateEdge] ($(p7.south)$);
  \draw ($(p11.north)$) edge[stateEdge] ($(p8.south)$);
  \draw ($(p11.north)$) edge[stateEdge] ($(p7.south)$);
  \draw ($(p10.north)$) edge[stateEdge] ($(p7.south)$);
  \draw ($(p12.north)$) edge[stateEdge] ($(p9.south)$);
  \draw ($(p12.north)$) edge[stateEdge] ($(p10.south)$);
  \draw ($(p14.north)$) edge[stateEdge] ($(p11.south)$);
  \draw ($(p14.north)$) edge[stateEdge] ($(p10.south)$);
  \draw ($(p15.north)$) edge[stateEdge] ($(p12.south)$);
  \draw ($(p15.north)$) edge[stateEdge] ($(p10.south)$);
  \draw ($(p15.north)$) edge[stateEdge] ($(p14.south)$);
\end{tikzpicture}
\caption{Known relations among fragments of FPQ}\label{fig:overview}
\end{figure}

Finally, Figure \ref{fig:overview} provides the implication of the known results on RDF graphs for the general relations between some interesting fragments of FPQ where $\mathcal{L}_1 \to \mathcal{L}_2$ to denote that $\mathcal{L}_1$ is expressible in $\mathcal{L}_2$. Note that this paper does not discuss all fragments of FPQ such as RPQ($\wedge$) while the left fragments leave open.

\subsection{Expressiveness of property paths in FPQ}
In syntax, \emph{property paths} (PP) in SPARQL 1.1 are inductively defined as follows \cite{sparql1.1}.
\begin{compactitem}
\item Any IRI in $I$ is a property path.
\item If $\elt_1$ and $\elt_2$ are property paths, then so are the followings: $\elt_1\, /\, \elt_2$ and $\elt_1\mid \elt_2$.
\item If $\elt$ is a property path, then so are the followings: $\elt?$, $\elt \ast$, $\elt +$, and $\hat{}\, \elt$.
\item If $\iri_i \in I$ for $1 \le i \le n+m$, then $!\elt$ is a property path where $\elt = (\iri_1 \mid \ldots \mid \iri_n \mid \hat{}\,\iri_{n+1} \mid \ldots \mid \hat{}\,\iri_{n+m})$.
\end{compactitem}

Semantically, let $P$ be a property path pattern of the form $(u, \elt, v)$ where $\elt$ is a property path, then the evaluation of $P$ over an RDF graph $G$ is defined as follows:
\begin{multline*}
\semm {P}{G} := \{\mu\mid \dom{\mu} = \var {\{u, v\}} \text{ and }\\ (\mu(u), \mu(v)) \in \semm {\elt}{G}\},
\end{multline*}
where $\semm {\elt}{G}$ is inductively defined as follows:
\begin{compactitem}
\item
$\semm {\iri}{G} := \{(a, b)\mid (a, \iri, b) \in G\}$.
\item
$\semm {\elt_1\,/\, \elt_2}{G} := \{(a, b) \mid \text{ there exists } c \text{ such that } (a, c) \in \semm {\elt_1}{G} \text{ and } (c, b) \in \semm {\elt_2}{G}\}$.
\item
$\semm {\elt_1\mid \elt_2}{G} := \semm {\elt_1}{G} \cup \semm {\elt_2}{G}$.
\item
$\semm {!(\iri_1 \mid \ldots \mid \iri_n)}{G} := \{(a, b)\mid  \exists \, c, (a, c, b) \in G \text{ and }$\\  $\forall\, i \in \{1, \ldots, n\}, (a, b) \not \in  \semm{\iri_i}{G}\}$.
\item
$\semm {\,\hat{}\,\elt}{G} := \{(a, b) \mid (b, a) \in \semm {\elt}{G}\}$.
\item
$\semm {\elt?}{G} := \{(a, a) \mid a\in \adom {G}\} \cup \semm {\elt}{G}$.
\item
$\semm {\elt+}{G} := \semm {\elt}{G} \cup \semm {\elt\,/\, \elt}{G} \cup \semm {\elt\,/\, \elt\, /\, \elt}{G} \cup \cdots$.
\item
$\semm {\elt\ast}{G} := \{(a, a) \mid a\in \adom {G}\} \cup \semm {\elt}{G} \cup \semm {\elt\,/\, \elt}{G} \cup \semm {\elt\,/\, \elt\, /\, \elt}{G} \cup \cdots$.
\end{compactitem}

A \emph{PP query} is of the form $(?x, \elt, ?y)$ where $\elt$ is a PP. Let $G$ be an RDF graph. $\semm {(?x, pp, ?y)} = \{(?x \to a, ?y \to b) \mid (a, b)\in \semm {pp}{G}\}$. For simplification, we still use PP to denote the PP query language, where each query is a PP query.

Since nre is not expressible in PP \cite{Taski_SPARQL}, we directly conclude the following proposition.
\begin{proposition}\label{prop:RPQ-not-PP}
NRPQ is not expressible in PP.
\end{proposition}

Moreover, since PP allows the negation of (atomic) property, PP is not expressible in nre \cite{Taski_SPARQL}. To prove that PP is not expressible in UFCNRPQ, we first introduce the following property named \emph{monotonicity}.

A UFCNRPQ query $\mathbf{q}$ is \emph{monotone} if for any two datasets $\mathbb{D} = (G, \mathcal{D})$ and $\mathbb{D}' = (G', \mathcal{D}')$, $\mathbb{D}\subseteq \mathbb{D}'$ implies $\semm {\mathbf{q}}{\mathbb{D}} \subseteq \semm {\mathbf{q}}{\mathbb{D}'}$. Here $\mathbb{D} \subseteq \mathbb{D}'$ is defined as follows:
\begin{compactitem}
\item $G \subseteq G'$;
\item for any $D\in \mathcal{D}$, there exists some $D' \in \mathcal{D}'$ such that $D \subseteq D'$.
\end{compactitem}

Since each UFCNRPQ query can be rewritten a conjunctive first-order query (CQ) which is monotone \cite{abs_book}, we conclude the following result.
\begin{lemma}\label{lem:PP-not-UFCRPQ}
All UFCNRPQ queries are monotone.
\end{lemma}

\begin{proposition}\label{prop:PP-not-UFCRPQ}
PP is not expressible in UFCNRPQ.
\end{proposition}

Since the negation-free PP can be expressible in nre \cite{Taski_SPARQL}, it is clear that the negation-free PP queries are also expressible in NRPQ. Theoretically, it is feasible to introduce nre with negation \cite{Taski_SPARQL} to extend our proposed FPQ.

\subsection{Expressiveness of FPQ in SPARQL}
To compare FPQ with SPARQL in expressiveness, we recall briefly nSPARQL \cite{nsparql}

In syntax, nSPARQL (graph) patterns is defined in an inductive way:
\begin{compactitem}
	\item Each nre-triples are nSPARQL patterns;
	\item $P_1\, \UNION\, P_2$, $P_1\, \andd\, P_2$, and $P_1\, \OPT\, P_2$ are nSPARQL patterns if $P_1$ and $P_2$ are patterns;
    \item $\SELECT_{S}(P)$ is an nSPARQL pattern if $P$ is an nSPARQL pattern and $S \subseteq V$;
	\item $P_1\, \FILTER\, C$ is an nSPARQL pattern if $P_1$ is an nSPARQL pattern and $C$ is a constraint.
\end{compactitem}

Semantically, the evaluation of general nSPARQL patterns is defined as follows:
\begin{compactitem}
\item $\semm {P_1 \UNION P_2}{G} = \semm {P_1} G \cup \semm {P_2} G.$
\item $\semm {P_1 \andd P_2}{G} = \semm {P_1} G \Join \semm {P_2} G$,
\item $\semm {P_1\, \OPT\, P_2} G = (\semm {P_1} G \Join \semm {P_2} G) \cup (\semm {P_1} G  \smallsetminus \semm {P_2} G)$,
 where $
\Omega_1 \smallsetminus \Omega_2 =
\{ \mu_1 \in \Omega_1 \mid \neg \exists \mu_2 \in \Omega_2 :
\mu_1 \sim \mu_2\}
$ for any two sets of mappings $\Omega_1$ and $\Omega_2$.
\item $\semm {\SELECT_{S}(P_{1})}{G} =\{\mu|_{S \cap \dom \mu} \mid \mu \in \semm{P_1}{G}\}$ where $f|_X$ is the restriction of a function $f$ to a subset $X$ of its domain.
\item $\semm {P_1 \FILTER C} G = \{\mu \in \semm {P_1} G \mid \mu(C) = \true \}.
$ Here, for any mapping $\mu$ and constraint $C$, the evaluation of
$C$ on $\mu$, denoted by $\mu(C)$, is defined in terms of a
three-valued logic with truth values $\true$, $\false$, and $\error$. Here we delete the semantics of filter conditions. For more details, please read some references \cite{perez_sparql_tods}.
\end{compactitem}

Since the Kleene star $\ast$ is not expressible in SPARQL \cite{nsparql}, let nre$^{\mathrm{sf}}$ be the Kleene star-free nre. We use RPQ$^{\mathrm{sf}}$ to denote RPQ by only allowing nre$^{\mathrm{sf}}$-expressions.

Since nSPARQL does not support querying on relations, we conclude the inexpressivity of  RPQ$^{\mathrm{sf}}(\mathrm{\mathsmaller R})$ in nSPARQL.
\begin{proposition}\label{prop:RPQ-not-nSPARQL}
RPQ$^{\mathrm{sf}}(\mathrm{\mathsmaller R})$ is not expressible in {nSPARQL}.
\end{proposition}

We use nSPARQL$^{\mathrm{sf}}$ to denote an extension of SPARQL by allowing the Kleene star-free nre$^{\mathrm{sf}}$-triple patterns.
\begin{proposition}\label{prop:RPQ-not-nSPARQL-sf}
RPQ$^{\mathrm{sf}}(|,\mathrm{\mathsmaller N},\wedge,\vee)$ is expressible in {nSPARQL$^{\mathrm{sf}}$}.
\end{proposition}

\begin{theorem}\label{thm:FPQ-SPARQL}
The following properties hold:
\begin{compactitem}
\item RPQ$^{\mathrm{sf}}(\mathrm{\mathsmaller R})$ is not expressible in SPARQL.
\item RPQ$^{\mathrm{sf}}(|,\mathrm{\mathsmaller N},\wedge,\vee)$ is expressible in SPARQL.
\end{compactitem}
\end{theorem}
%

In short, the Kleene star $\ast$ in nre and the federated operator $\mathrm{\mathsmaller R}$ are indeed beyond the expressiveness of SPARQL.

At the end of this section, we will discuss the complexity of the query evaluation problem in FPQ.

Let $\mathbb{D} = (G, \mathcal{D})$ be a heterogeneous database. Given a FCNRPQ $\mathbf{q}(u, v)$ and a mapping $\mu$, the \emph{query evaluation problem} is deciding whether $\mu \in \semm {\mathbf{q}(u, v)}{G}$, that is, whether the tuple $\mu$ is in the result of the query $\mathbf{q}$ on the heterogeneous database $\mathbb{D}$.

There are two kinds of computational complexity in the query evaluation problem \cite{abs_book,ahv_book}:
\begin{compactitem}
\item the \emph{data complexity} refers to the complexity w.r.t. the size of the heterogeneous database $\mathbb{D}$, given a fixed query $\mathbf{q}$; and
\item the \emph{combined complexity} refers to the complexity w.r.t. the size of query $\mathbf{q}$ and the heterogeneous database $\mathbb{D}$.
\end{compactitem}

As a result, we can conclude the following proposition.
\begin{proposition}\label{prop:FCNRPQ_complexity}
The followings hold:
\begin{enumerate}
\item The data complexity of the query evaluation of FCNRPQ is in polynomial time;
\item The combined complexity of the query evaluation of FCNRPQ is in NP-complete time.
\end{enumerate}
\end{proposition}

Note that the query evaluation of UFCNRPQ has the same complexity as
the evaluating of FCNRPQ since we can simply evaluate a number (linear in the
size of a FCNRPQ) of FCNRPQ in isolation \cite{ahv_book}.

\section{Experiments}\label{sec:implement}
\begin{figure}[H]
  \centering
  \includegraphics[width=0.5\textwidth]{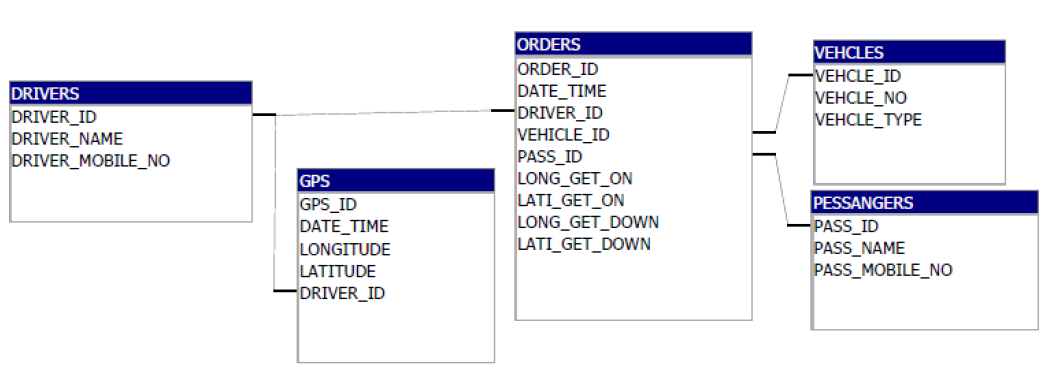}\\
  \caption{Relational database.}\label{fig:database}
\end{figure}

All experiments are carried out on a machine with operating system WINDOWS 7 (professional version) having following specifications like CPU with four cores of 3.30GHz,4GB memory and 450 GB storage. MySQL is used as relational database tool. Our code is an extension of RPL \cite{RPL} which evaluates RPQs on RDF graphs \cite{Regular Path Queries on Large Graphs}. Firstly we construct a relational database as in Figure \ref{fig:database}.

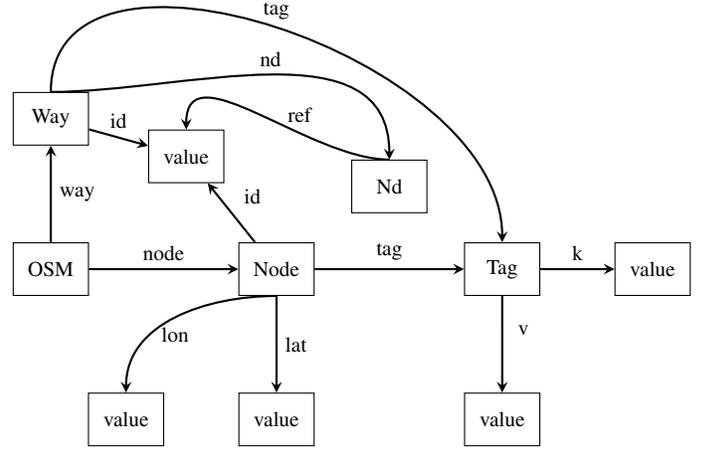
\begin{figure}\centering
    \tikzstyle{entity} = [rectangle, minimum width=1cm, minimum height=0.7cm, text centered, font=\small, draw=black]
    \tikzstyle{final} = [ellipse, minimum width=1.5cm, minimum height=0.7cm, text centered, text = black, font=\small, draw=black]
    \tikzstyle{arrow} = [thick,->,>=stealth, font=\small]

    \begin{tikzpicture}[scale=0.35][node distance=1.5cm]
        \node(entity1)[entity]{OSM};
        \node(entity2)[entity,right of=entity1,xshift=2cm]{Node};
        \node(entity3)[entity,above of=entity1,yshift=1cm]{Way};
        \node(entity4)[entity,right of=entity2,xshift=2cm]{Tag};
        \node(entity5)[entity,right of=entity4,xshift=1cm]{value};
        \node(entity6)[entity,below of=entity4,yshift=-1cm]{value};
        \node(entity7)[entity,above of=entity2,yshift=0.5cm,xshift=-1.2cm]{value};
        \node(entity10)[entity,above of=entity2,yshift=0.1cm,xshift=1.5cm]{Nd};
        \node(entity8)[entity,below of=entity2,xshift=-2cm,yshift=-1cm]{value};
        \node(entity9)[entity,below of=entity2,yshift=-1cm]{value};

        \draw[arrow](entity1) -- node[anchor=west]{way}(entity3);
        \draw[arrow](entity1) -- node[anchor=south]{node}(entity2);
        \draw[arrow](entity2) -- node[anchor=south]{tag}(entity4);
        \draw[arrow](entity4) -- node[anchor=south]{k}(entity5);
        \draw[arrow](entity4) -- node[anchor=south]{\qquad v}(entity6);
        \draw[arrow](entity2) -- node[anchor=south]{\qquad id}(entity7);
        \draw[arrow](entity2.south) to[out=180,in=90] node[anchor=north]{lon} (entity8.north);
        \draw[arrow](entity2) -- node[anchor=west]{lat}(entity9);
        \draw[arrow](entity3) -- node[anchor=south]{id}(entity7);
        \draw[arrow](entity3.north) to[out=90,in=90] node[anchor=south]{tag} (entity4.north);
        \draw[arrow](entity3.north) to[out=0,in=90] node[anchor=south]{nd} (entity10.north);
        \draw[arrow](entity10.north)to[out=180,in=90]node[anchor=west]{\quad ref}(entity7.north);
    \end{tikzpicture}
\caption{RDF Graph.}\label{fig:rdf}
\end{figure}

We assessed our federated path queries on relational databases and a real RDF data set of total size of 14000 lines. It provides the information about a map in longitude and latitude points for different locations as in Figure \ref{fig:rdf}.

Four federated queries are planned for experiments these are as followed:

\paragraph{Query 1} On specific Date, at what location passengers get on vehicle and get off.

Let $\mathbf{q}_1$ be an FPQ query defined as follows:
\begin{equation*}
\mathbf{q}_1(?x, ?y) = (?x, exp_1, ?y) \wedge R(\bar{u}),
\end{equation*}
where
\begin{compactitem}
\item $exp_1: \nexts^{-1}::\emph{lon}/\nexts::\emph{lat}$;
\item $R(\bar{u}) := Orders(\emph{Date}, ?x, ?y)$.
\end{compactitem}

The $exp_1$ is to query the latitude and longitude of the points in the map. The $R(\bar{u})$ is to query the relational database on specific date for which location \emph{Longitude and latitude points} the passengers have place the order. Finally, in $\mathbf{q}_5$ the parts are joined in a federation.

\begin{table*}[htbp]
\centering
\caption{Query performance on different sizes of relational databases}\label{tab:query-performance}
\small
\begin{tabular}{|c|c|c|c|c|c|c|c|c|c|c|c|c|c|} \hline
\multicolumn{2}{|c|}{Query} & \multicolumn{12}{|c|}{Relational Database} \\ \cline{3-14}
\multicolumn{2}{|c|}{} & \multicolumn{2}{|c|}{1,000,000 tuples} & \multicolumn{2}{|c|}{2,000,000 tuples} & \multicolumn{2}{|c|}{3,000,000 tuples} & \multicolumn{2}{|c|}{4,000,000 tuples} & \multicolumn{2}{|c|}{5,000,000 tuples} & \multicolumn{2}{|c|}{6,400,000 tuples} \\ \cline{3-14}
\multicolumn{2}{|c|}{}  & Time & Solutions &Time & Solutions & Time & Solutions & Time & Solutions &Time & Solutions &Time & Solutions  \\ \hline
 & RDF & 2,122 & 2,330 & 1,985 & 2,330 & 2,040 & 2,330 & 2,028 & 2,330 & 2,023 & 2,330 & 2,341 & 2,330 \\  \cline{2-14}
  $\mathbf{q}_1$  & Rel-DB & 926 & 77,179 & 1,686 & 82,497  & 2,431 &  82,548& 2,883 & 82,548 &4,038 & 82,771 & 6,099  & 102,207  \\  \cline{2-14}
 & Joining & 728 & 13,920 & 791 & 14,798 & 782 & 14,801 & 774 & 14,801 & 780 & 14,825 & 1,016 &  17,857 \\  \cline{2-14}
 & Total & 3,778 & 13,920  & 4,462  &14,798  &5,254  & 14,801  & 5,685 & 14,801 & 6,842 & 14,825 & 9,456 & 17,857 \\ \hline
 & RDF &18,217  &15  &18,679  &15  &18,736  &15  &18,489  &15  &18,620  &15  &21,147  &15  \\  \cline{2-14}
  $\mathbf{q}_2$  & Rel-DB &952  &77,179  &1,612  &82,497  &2,307  &82,548  &2,893  &82,548  &4,135  &82,771  &6,494  &102,207  \\  \cline{2-14}
 & Joining &16  &1,815  &16  &1964  &16  &1,967  &16  &1,967  &16  &1,970  &26  &2,411  \\  \cline{2-14}
 & Total &19,186  &1,815  &20,309  &1,964  &21,060  &1,967  &21,398  &1,967  &22,771  &1,970  &27,667  &2,411  \\ \hline
 & RDF &18,393  &25  &18,394  &25  &18,470  &25  &18,755  &25  &18,739  &25  &19,701  &25  \\  \cline{2-14}
  $\mathbf{q}_3$  & Rel-DB &961  &77,179  &1,751  &82,497  &2,468  &82,548  &2,867  &82,548  &3,930  &82,771  &6,618  &102,207  \\  \cline{2-14}
 & Joining &18  &3,280  &19  &3,465  &18  &3,468  &18  &3,468  & 18 &3,471  &33  &4,291  \\  \cline{2-14}
 & Total &19,373  &3,280  &20,166  &3,465  &20,975  &3,468  &21,640  &3,468  &22,688  &3,471  &26,352  &4,291  \\ \hline
 & RDF &18,530  &14  &18,628  &14  &18,616  &14 &18,910  &14  &19,106  &14  &19,223  &14   \\  \cline{2-14}
  $\mathbf{q}_4$  & Rel-DB &1,512  &15  &3,303  &15  &4,534  &15  &6,770  &15  &6,845  &16  &12,144  &16  \\  \cline{2-14}
 & Joining &0.3  &5  &0  &5  &0  &5  &0  &5  &0  &5  &0  &5  \\  \cline{2-14}
 & Total &20,043  &5  &21,931  &5  &23,151  &5  &25,681  &5  &25,951  &5  &31,367  &5  \\ \hline
\end{tabular}
\end{table*}

\paragraph{Query 2} On specific Date, Did the passengers visit a tourist attraction place on a map?

Let $\mathbf{q}_2$ be an FPQ query defined as follows:
\begin{equation*}
\mathbf{q}_2(?x, ?y) = (?x, exp_2, ?y) \wedge R(\bar{u}),
\end{equation*}
where
\begin{compactitem}
\item $exp_2: [\nexts^{-1}::\emph{lon}/[\nexts::\emph{tag}/\edge::\emph{`tourism'}]]/$\\
$[[\nexts::\emph{tag}/\edge::\emph{`tourism'}]/\nexts::\emph{lat}]$;
\item $R(\bar{u}) := Orders(\emph{Date}, ?x, ?y)$.
\end{compactitem}

The $exp_2$ is to query the latitude and longitude of the tourist attraction points in the map. The $R(\bar{u})$ is to query the relational database which points that the passengers have gone through the order. Finally, by joining the parts we get $Q_2$.

\paragraph{Query 3} On specific and unique Date, on which location of the road passengers get down from the taxi.

Let $\mathbf{q}_3$ be an FPQ query defined as follows:
\begin{equation*}
\mathbf{q}_3(?x, ?y) = (?x, exp_1, ?y) \wedge (?x, exp_3, ?y) \wedge R(\bar{u}),
\end{equation*}
where
\begin{compactitem}
\item $exp_1: \nexts^{-1}::\emph{lon}/\nexts::\emph{lat}$;
\item $exp_3: \nexts^{-1}::\emph{lon}/[\self::[\nexts^{-1}::\emph{ref}]]/\nexts::\emph{lat}$;
\item $R(\bar{u}) := Orders(\emph{Date}, ?x, ?y)$.
\end{compactitem}

The $exp_1$ is the same as above in $\mathbf{q}_1$. The $exp_3$ is to query the latitude and longitude of points on the road in RDF data set. Sometimes some points are not on the road. The $R(\bar{u})$ is to query that on exact date to which points that the passengers have placed the order. Finally, by joining three parts we have the result.

\paragraph{Query 4} On specific Date, Can a passenger take a ride when No vehicle is available at stand. Can already on the way driver
accommodate another passenger on his way by choosing the right path?

Let $\mathbf{q}_4$ be an FPQ query defined as follows:
\begin{multline*}
\mathbf{q}_4(?x, ?y) = (?x, exp_4, ?y) \wedge (?x, exp_5, ?y) \wedge (?x, exp_6, ?y) \\ \wedge [R(\bar{v}) \wedge R(\bar{e})],
\end{multline*}
where
\begin{compactitem}
\item $exp_4: \nexts^{-1}::\emph{lon}$;
\item $exp_5: \nexts::\emph{lat}$;
\item $exp_6: \nexts::nd/\nexts::ref/\nexts^{-1}::\emph{id}$;
\item $R(\bar{v} := GPS(\emph{Date}, \emph{lon}, \emph{lat}, ?driver id)$;
\item $R(\bar{u}) := Orders(\emph{Date}, ?lon, ?lat, ?driver id)$.
\end{compactitem}

The $exp_4$ gets the longitude of points in the map. The $exp_5$ gets the latitude of points in the map. The $exp_6$ gets the
information of ways. The $[R(\bar{v}) \wedge R(\bar{e})]$ gets the order information that matches the current time and starting point. Finally, by joining the four parts in return we have the result whether a driver can respond to another passenger and accommodate
him/her by choosing the right path.

For the first three queries, from the relational database, we got the information about at which points the passengers get in and out of the vehicle, and for confirming that points are tourist attraction location or on the road for that kind of information we need to use map (RDF dataset). For the Query Four, from the relational database, we can get the information  about the detials of orders. but for the confirmation about  that a driver  already with a passanger or on the way but not with passenger can fulfil the order placed by another passenger. The answer is yes, It can accomodate another passenger as we have the information about the path and location point of the passanger by querying the RDF dataset in fedration of reltional database. The path descovery and information about the right location of travller  becomes possible by using the federated queries like $Q_1$, $Q_2$, $Q_3$, and $Q_4$. Second part of experiment is to test the performance of these queries for different sizes of Datasets. For that we  found following results  shown  above in the Table \ref{tab:query-performance}. Their graphical presentation is presented at end of the document. Which is show the performance of the above said  fedrated queries and thier comparison.we found  the fedrated path queries are more effective in finding required results at comparitivly less computation power.

\section{Conclusions}\label{sec:conclusion}
We have proposed federated path queries to navigate through RDF graphs integrated with relational databases. Some investigation about some fundamental properties of those federated path queries. We prove that FPQ strictly expresses nested regular expression and we also give a complete Hasse diagram of fragments of FPQ. Finally, we show that the query evaluation of FPQ maintains the polynomial time data complexity and NP-complete combined complexity as the same as conjunctive first-order queries. These results provides a starting point for further research on expressiveness of federated path languages for heterogeneous databases such as RDF graphs integrated with relational databases. Besides, we show that federated path queries can be evaluated efficiently in our experiments.

There are a number of practical open problems like more complex queries on larger heterogeneous datasets of database, to formulate relationships between within heterogeneous RDFs and with heterogeneous relational databases in different scenarios ultimately toward an optimized query manager. In this paper, we restrict that RDF data does not contain \emph{blank nodes} as the same treatment in
nSPARQL. We have to admit that blank nodes do make RDF data more expressive
since a blank node in RDF is taken as an existentially quantified variable
\cite{DBLP:journals/ws/HoganAMP14}.  An interesting future work is to extend
our proposed federated path queries for general RDF data with blank nodes by allowing path variables which are already valid in some extensions of SPARQL such as SPARQLeR\cite{DBLP:conf/esws/KochutJ07} and CPSPARQL
\cite{DBLP:journals/ijwis/AlkhateebE14}, which
are popular in querying over general RDF data with blank nodes.

\section{Acknowledgments}
This work is supported by the programs of the National Key Research and Development Program of China (2016YFB1000603), the National Natural Science Foundation of China (NSFC) (61502336), and the open funding project of Key Laboratory of Computer Network and Information Integration (Southeast University), Ministry of Education (K93-9-2016-05). Xiaowang Zhang is supported by Tianjin Thousand Young Talents Program.

%
%
%

\onecolumn

\begin{figure}[h]
\begin{minipage}{0.48\linewidth}
\begin{tikzpicture}
\begin{axis}[
    title={Query 1},
    xlabel={Tuples in relational database[in millions]},
    ylabel={Time[ms]},
    symbolic x coords={1,2,3,4,5,6.4},
    legend pos=north west,
    ymajorgrids=true,
    grid style=dashed,
]

\addplot[
    color=black,
    mark=triangle*, mark options={fill=white}
    ]
    coordinates {
    (1,2123)(2,1985)(3,2040)(4,2028)(5,2023)(6.4,2341)
    };

\addplot[
    color=orange,
    mark=*, mark options={fill=blue}
    ]
    coordinates {
    (1,927)(2,1686)(3,2486)(4,2883)(5,4038)(6.4,6099)
    };
\addplot[
    color=blue,
    mark=square*, mark options={fill=red}
    ]
    coordinates {
    (1,728)(2,791)(3,782)(4,774)(5,780)(6.4,1016)
    };

\addplot[
    color=blue,
    mark=square*, mark options={fill=black}
    ]
    coordinates {
    (1,3778)(2,4462)(3,5254)(4,5685)(5,6842)(6.4,9456)
    };
    \legend{RDF,Rel-Data,Joining,Total}
\end{axis}

\hskip 9cm

\begin{axis}[
    title={Query 2},
    xlabel={Tuples in relational database[in millions]},
    ylabel={Time[ms]},
    symbolic x coords={1,2,3,4,5,6.4},
    legend style={at={(0.35,0.55)}},  
    ymajorgrids=true,
    grid style=dashed,
]

\addplot[
    color=black,
    mark=triangle*, mark options={fill=white}
    ]
    coordinates {
    (1,18217)(2,18680)(3,18737)(4,18489)(5,18620)(6.4,21147)
    };
\addplot[
    color=orange,
    mark=*, mark options={fill=blue}
    ]
    coordinates {
    (1,953)(2,1613)(3,2308)(4,2893)(5,4135)(6.4,6494)
    };

\addplot[
    color=blue,
    mark=square*, mark options={fill=red}
    ]
    coordinates {
    (1,16)(2,17)(3,16)(4,16)(5,16)(6.4,26)
    };

\addplot[
    color=blue,
    mark=square*, mark options={fill=black}
    ]
    coordinates {
    (1,19186)(2,20309)(3,21060)(4,21398)(5,22771)(6.4,27667)
    };

    \legend{RDF,Rel-Data,Joining,Total}

\end{axis}
\end{tikzpicture}
\end{minipage}

\begin{minipage}{0.48\linewidth}
\begin{tikzpicture}
\begin{axis}[
    title={Query 3},
    xlabel={Tuples in relational database[in millions]},
    ylabel={Time[ms]},
    symbolic x coords={1,2,3,4,5,6.4},
    legend style={at={(0.35,0.55)}},  
    ymajorgrids=true,
    grid style=dashed,
]

\addplot[
    color=black,
    mark=triangle*, mark options={fill=white}
    ]
    coordinates {
    (1,18393)(2,18395)(3,18471)(4,18755)(5,18739)(6.4,19701)
    };
\addplot[
    color=orange,
    mark=*, mark options={fill=blue}
    ]
    coordinates {
    (1,962)(2,1752)(3,2486)(4,2867)(5,3930)(6.4,6618)
    };

\addplot[
    color=blue,
    mark=square*, mark options={fill=red}
    ]
    coordinates {
    (1,18)(2,19)(3,18)(4,18)(5,19)(6.4,33)
    };
\addplot[
    color=blue,
    mark=square*, mark options={fill=black}
    ]
    coordinates {
    (1,19373)(2,20166)(3,20975)(4,21640)(5,22688)(6.4,26352)
    };

    \legend{RDF,Rel-Data,Joining,Total}

\end{axis}
\hskip 9cm
\begin{axis}[
    title={Query 4},
    xlabel={Tuples in relational database[in millions]},
    ylabel={Time[ms]},
    symbolic x coords={1,2,3,4,5,6.4},
    legend style={at={(0.35,0.55)}},  
    ymajorgrids=true,
    grid style=dashed,
]

\addplot[
    color=black,
    mark=triangle*, mark options={fill=white}
    ]
    coordinates {
    (1,18530)(2,18628)(3,18616)(4,18910)(5,19106)(6.4,19223)
    };
\addplot[
    color=orange,
    mark=*, mark options={fill=blue}
    ]
    coordinates {
    (1,1513)(2,3303)(3,4534)(4,6770)(5,6845)(6.4,12144)
    };

\addplot[
    color=blue,
    mark=square*, mark options={fill=red}
    ]
    coordinates {
    (1,0)(2,0)(3,0)(4,0)(5,0)(6.4,0)
    };

\addplot[
    color=blue,
    mark=square*, mark options={fill=black}
    ]
    coordinates {
    (1,20043)(2,21931)(3,23151)(4,25681)(5,25951)(6.4,31367)
    };

    \legend{RDF,Rel-Data,Joining,Total}
\end{axis}
\end{tikzpicture}
\end{minipage}
\end{figure}
\begin{figure}[H]
\begin{center}
\begin{tikzpicture}
\begin{axis}[
    title={Query time},
    xlabel={Tuples in relational database[in millions]},
    ylabel={Time[ms]},
    symbolic x coords={1,2,3,4,5,6.4},
    legend style={at={(0.35,0.5)}},  
    ymajorgrids=true,
    grid style=dashed,
]

\addplot[
    color=blue,
    mark=square*, mark options={fill=black}
    ]
    coordinates {
    (1,3778)(2,4462)(3,5254)(4,5685)(5,6842)(6.4,9456)
    };

\addplot[
    color=blue,
    mark=square*, mark options={fill=red}
    ]
    coordinates {
    (1,19186)(2,20309)(3,21060)(4,21398)(5,22771)(6.4,27667)
    };

\addplot[
    color=black,
    mark=triangle*, mark options={fill=white}
    ]
    coordinates {
    (1,19373)(2,20166)(3,20975)(4,21640)(5,22688)(6.4,26352)
    };
\addplot[
    color=orange,
    mark=*, mark options={fill=blue}
    ]
    coordinates {
    (1,20043)(2,21931)(3,23151)(4,25681)(5,25951)(6.4,31367)
    };
    \legend{Query-1,Query-2,Query-3,Query-4}

\end{axis}
\end{tikzpicture}
\end{center}
\end{figure}\label{fig:queryTime}
\twocolumn

\newpage 
\section*{Appendix: Proofs}

\noindent Proof of \textbf{Lemma} \ref{lem:nre-plus}\\
By induction on the structure of $e$.
\begin{compactitem}	
\item	If $e$ is of the form $\axis$ or $\axis^{-1}$ and $(a, b) \in \semm {e}{G}$ for some \emph{p}-RDF graph $G$ and some pair $(a, b)$ with $a, b \in \const{G}$ then let consider seven cases of $\axis$ as follows:
	\begin{compactitem}
		\item If $\axis$ is $\self$ then $G \neq \emptyset$. Let $H = \{(a, p, b) \mid (a, p, b) \in G\}$ be a strongly acyclic \emph{p}-RDF graph, $\semm {e}{H} \neq \emptyset$.
		\item If $\axis$ is $\nexts$ then $(a, p, b) \in G$. Let $H = \{(a, p, b)\}$ be a strongly acyclic \emph{p}-RDF graph, $\semm {e}{H} \neq \emptyset$.
		\item If $\axis$ is $\nexts^{-1}$ then $(b, p, a) \in G$. Let $H = \{(b, p, a)\}$ be a strongly acyclic \emph{p}-RDF graph, $\semm {e}{H} \neq \emptyset$.
		\item If $\axis$ is $\edges$ then $(a, b, c) \in G$ (in this case $b = p$). Let $H = \{(a, b, c)\}$ be a strongly acyclic \emph{p}-RDF graph, $\semm {e}{H} \neq \emptyset$.
		\item If $\axis$ is $\edges^{-1}$ then $(b, a, c) \in G$ (in this case $a = p$). Let $H = \{(b, a, c)\}$ be a strongly acyclic \emph{p}-RDF graph, $\semm {e}{H} \neq \emptyset$.
		\item If $\axis$ is $\nodes$ then $(c, a, b) \in G$ (in this case $a = p$). Let $H = \{(c, a, b)\}$ be a strongly acyclic \emph{p}-RDF graph, $\semm {e}{H} \neq \emptyset$.
		\item If $\axis$ is $\edges^{-1}$ then $(c, b, a) \in G$ (in this case $a = p$). Let $H = \{(c, b, a)\}$ be a strongly acyclic \emph{p}-RDF graph, $\semm {e}{H} \neq \emptyset$.
	\end{compactitem}
\item If $e$ is of the form $e_1|e_2$ then this claim readily holds by induction.
\item If $e$ is of the form $e_1/e_2$ and $(a, b) \in \semm {e}{G}$ for some \emph{p}-RDF graph $G$ then there exists some $c\in \const{G}$ such that $(a, c) \in \semm {e_1}{G}$ and $(c, b) \in \semm {e_1}{G}$. By induction, let $H_i$ be a strongly acyclic \emph{p}-RDF graph for $i=1,2$, $(a, c) \in \semm {e_1}{H_1}$ and $(c, b)\semm {e_2}{H_2}$. By renaming, we can obtain two new strongly acyclic \emph{p}-RDF graphs $H'_1$ and $H'_2$ respectively with $\const{{H'_1}} \cap \const{{H'_2}} = \{p, c\}$. Let us construct a \emph{p}-RDF graph $H$ by the union of $H'_1$ and $H'_2$. Clearly, $H$ is strongly acyclic by our assumption. Thus $(a, c) \in \semm {e_1}{H'_1}$ and $(c, b) \in \semm {e_2}{H'_2}$. Moreover, $(a, c) \in \semm {e_1}{H}$ and $(c, b) \in \semm {e_2}{H}$ because of the monotonicity of nre$^{\mathrm{cf}}$ and $H'_1 \subseteq H$ and $H'_2 \subseteq H$. Therefore, $(a, b) \in \semm {e}{H}$.
\item If $e$ is of the form $e^{\ast}$ then this claim readily holds by using two cases of $e_1|e_2$ and $e_1/e_2$.
\item Finally, if $e$ is of the form $\axis::[e_1]$ and $(a, b) \in \semm {e}{G}$ for some \emph{p}-RDF graph $G$ then let us consider seven cases.
	\begin{compactitem}
		\item If $e$ is of the form $\self::[e_1]$ (in this case $a = b$) then there exists some $c \in \const{G}$ such that $(b, c) \in \semm {e_1}{G}$. By induction, let $H_1$ be a strongly acyclic \emph{p}-RDF graph, $(b, c) \in \semm {e_1}{H_1}$. Thus $(a, b) \in \semm {\self::[e_1]}{H_1}$. Therefore $G_1$ is desired.
		\item If $e$ is of the form $\nexts::[e_1]$ then $(a, p, b) \in G$ and there exists some $c \in \const{G}$ such that $(p, c) \in \semm {e_1}{G}$. By induction, let $H_1$ be a strongly acyclic \emph{p}-RDF graph, $(p, c) \in \semm {e_1}{H_1}$. By renaming, we can obtain an RDF graph $H'_1$ by renaming such that $\{a, b\} \cap \const{H'_1} = \emptyset$.  Clearly,  $H'_1$ be a strongly acyclic \emph{p}-RDF graph. Let $H = H'_1 \cup \{(a, p, b)\}$. Moreover, $(p, c) \in \semm {e_1}{H}$  because of the monotonicity of nre$^{\mathrm{cf}}$ and $H'_1 \subseteq H$. Then $(a, b) \in \semm {\nexts::[e_1]}{H}$. Therefore $H$ is desired.
		\item If $e$ is of the form $\nexts^{-1}::[e_1]$ then $(b, p, a) \in G$ and there exists some $c \in \const{G}$ such that $(p, c) \in \semm {e_1}{G}$. By induction, let $H_1$ be a strongly acyclic \emph{p}-RDF graph, $(p, c) \in \semm {e_1}{H_1}$. By renaming, we can obtain an RDF graph $H'_1$ by renaming such that $\{a, b\} \cap \const{H'_1} = \emptyset$. Clearly,  $H'_1$ be a strongly acyclic \emph{p}-RDF graph. Let $H = H'_1 \cup \{(b, p, a)\}$. Moreover, $(p, c) \in \semm {e_1}{H}$  because of the monotonicity of nre$^{\mathrm{cf}}$ and $H'_1 \subseteq H$. Then $(a, b) \in \semm {\nexts^{-1}::[e_1]}{H}$. Therefore $H$ is desired.
		\item If $e$ is of the form $\edges::[e_1]$ then $(a, b, c) \in G$ (in this case $b = p$) and there exists some $d \in \const{G}$ such that $(c, d) \in \semm {e_1}{G}$. By induction, let $H_1$ be a strongly acyclic \emph{p}-RDF graph, $(c, d) \in \semm {e_1}{H_1}$. By renaming, we can obtain an RDF graph $H'_1$ by renaming such that $a\not \in \cap \const{H'_1} = \emptyset$.  Clearly,  $H'_1$ be a strongly acyclic \emph{p}-RDF graph. Let $H = H'_1 \cup \{(a, b, c)\}$. Moreover, $(c, d) \in \semm {e_1}{H}$  because of the monotonicity of nre$^{\mathrm{cf}}$ and $H'_1 \subseteq H$. Then $(a, b) \in \semm {\nexts::[e_1]}{H}$. Therefore $H$ is desired.
		\item If $e$ is of the form $\edges::[e_1]$ then $(b, a, c) \in G$ (in this case $a = p$) and there exists some $d \in \const{G}$ such that $(c, d) \in \semm {e_1}{G}$. By induction, let $H_1$ be a strongly acyclic \emph{p}-RDF graph, $(c, d) \in \semm {e_1}{H_1}$. By renaming, we can obtain an RDF graph $H'_1$ by renaming such that $a\not \in \cap \const{H'_1} = \emptyset$.  Clearly,  $H'_1$ be a strongly acyclic \emph{p}-RDF graph. Let $H = H'_1 \cup \{(b, a, c)\}$. Moreover, $(c, d) \in \semm {e_1}{H}$  because of the monotonicity of nre$^{\mathrm{cf}}$ and $H'_1 \subseteq H$. Then $(a, b) \in \semm {\nexts::[e_1]}{H}$. Therefore $H$ is desired.
		\item If $e$ is of the form $\nodes::[e_1]$ then $(c, a, b) \in G$ (in this case $a = p$) and there exists some $d \in \const{G}$ such that $(c, d) \in \semm {e_1}{G}$. By induction, let $H_1$ be a strongly acyclic \emph{p}-RDF graph, $(c, d) \in \semm {e_1}{H_1}$. By renaming, we can obtain an RDF graph $H'_1$ by renaming such that $a\not \in \cap \const{H'_1} = \emptyset$.  Clearly,  $H'_1$ be a strongly acyclic \emph{p}-RDF graph. Let $H = H'_1 \cup \{(c, a, b)\}$. Moreover, $(c, d) \in \semm {e_1}{H}$  because of the monotonicity of nre$^{\mathrm{cf}}$ and $H'_1 \subseteq H$. Then $(a, b) \in \semm {\nexts::[e_1]}{H}$. Therefore $H$ is desired.
		\item If $e$ is of the form $\nodes::[e_1]$ then $(c, b, a) \in G$ (in this case $b = p$) and there exists some $d \in \const{G}$ such that $(c, d)\semm {e_1}{G}$. By induction, let $H_1$ be a strongly acyclic \emph{p}-RDF graph, $(c, d) \in \semm {e_1}{H_1}$. By renaming, we can obtain an RDF graph $H'_1$ by renaming such that $a\not \in \cap \const{H'_1} = \emptyset$.  Clearly,  $H'_1$ be a strongly acyclic \emph{p}-RDF graph. Let $H = H'_1 \cup \{(c, b, a)\}$. Moreover, $(c, d) \in \semm {e_1}{H}$  because of the monotonicity of nre$^{\mathrm{cf}}$ and $H'_1 \subseteq H$. Then $(a, b) \in \semm {\nexts::[e_1]}{H}$. Therefore $H$ is desired.
	\end{compactitem}
	\end{compactitem}
Therefore, there always exists some strongly acyclic \emph{p}-RDF graph $H$ such that $(a, b) \in \semm {e}{H}$.

\noindent Proof of \textbf{Proposition} \ref{prop:intersection-nre}\\
Consider an nre($\cap$)-expression $e$ of the form $\nexts \cap \nexts/\nexts$. Suppose, for the sake of contradiction, that $e$ is expressible as $e'$ for some nre-expression $e$. Moreover, we can assume that $e'$ is constant-free. (Otherwise, assume that $c$ occurs in $e$, let us consider an RDF graph $G = \{(a, p, b), (b, p, d), (a, p, d)\}$ without $c$ and such that $\semm {e}{G} \neq \emptyset$. Since $e$ is expressible as $e'$, $\semm {e}{G} \neq \emptyset$. In other words, the constant $c$ can be removed from $e'$.

Consider an \emph{p}-RDF graph $G = \{(a, p, b), (b, p, c), (a, p, c)\}$, $(a, c) \in \semm {e}{G}$. Since $e$ is expressible as $e'$, we have $(a, c) \in \semm {e'}{G}$. By Lemma \ref{lem:nre-plus}, there exists some strongly acyclic \emph{p}-RDF graph $H$ such that $(a, c) \in \semm {e'}{H}$.
	
However, we can claim that
\begin{claim}
For any \emph{p}-RDF graph $G$, if $\semm {e}{G} \neq \emptyset$ then $\ind{G}$ is not strongly acyclic.
\end{claim}
Assume that $(a, c) \in \semm {e}{G}$. Then $G$ must contain some subgraph $G' = \{(a, p, b), (b, p, c), (a, p, c)\}$. However, $\ind{G'}$ is not strongly acyclic. Therefore, $\ind{G}$ is not strongly acyclic since $G' \subseteq G$.

\noindent Proof of \textbf{Theorem} \ref{thm:CNRPQ-NRPQ}\\
Let $G$ be an RDF graph. Consider a CNRPQ $\mathbf{q}(?x, ?y) $=$ (?x, e_1, ?y) $ $\wedge (?x, e_2, ?y)$. Suppose, for the sake of contradiction, that $\mathbf{q}(?x, ?y)$ is expressible as $\mathbf{q}'(?x, ?y)$ for some NRPQ $\mathbf{q}'(?x, ?y)$. Without loss of generality, we assume that $\mathbf{q}'(?x, ?y) = (?x, e, ?y)$ where $e$ is an nre-expression. Since $\mathbf{q}(?x, ?y)$ is expressible as $\mathbf{q}'(?x, ?y)$,
we have that $\semm {\mathbf{q}'(?x, ?y)}{G} = \semm {(?x, e, ?y)}{G} = \semm {\mathbf{q}(?x, ?y)}{G}$ $ = \semm {(?x, e_1, ?y) \wedge (?x, e_2, ?y)}{G} = \semm {(?x, e_1 \cap e_2, ?y)}{G}$. That is, $\semm {(?x, e, ?y)}{G} = \semm {(?x, e_1 \cap e_2, ?y)}{G}$. Therefore, $e_1\cap e_2$ is expressible as $e$, however, we have arrived a contradiction.

\noindent Proof of \textbf{Lemma} \ref{lem:ecnrpq-cnrpq}\\
By induction on the structure of $\mathbf{q}$.
\begin{compactitem}
\item If $\mathbf{q}(?x, ?y) = (?x, e, ?y)$ where $e$ is an nre-expression then it follows definitions.
\item If $\mathbf{q}(?x, ?y)$ is of the form (\ref{equ:CNRE}) then, by definition, we conclude that $\semm {\mathbf{q}(?x, ?y)}{G} =  \semm {(u_1, e_1, v_1)}{G} \Join \ldots \Join \semm {(u_n, e_n, v_n)}{G}$. Since for any $\mu_i\in \semm {(u_i, e_i, v_i)}{G}$ , the range of $\mu_i$ is $\{a\}$, $\mu_i \sim \mu_j$ for $i,j \in \{1,2,\ldots, n\}$. Then $\mu_1\cup \ldots \cup \mu_n \in \semm {(u_1, e_1, v_1)}{G} \Join \ldots \Join \semm {(u_n, e_n, v_n)}{G}$ $= \semm {\mathbf{q}(?x, ?y)}{G}$. So, $\semm {\mathbf{q}(?x, ?y)}{G}$ is not empty.
\end{compactitem}
Therefore, $\semm {\mathbf{q}(?x, ?y)}{G}\neq \emptyset$.

\noindent Proof of \textbf{Theorem} \ref{thm:ecnrpq-cnrpq}\\
Consider an FCNRPQ $\mathbf{q}(?x, ?y) = R(?x, ?y) \wedge (?x, \nexts, ?y)$. Suppose, for the sake of contradiction, that $\mathbf{q}(?x, ?y)$ is expressible as $\mathbf{q}'(?x, ?y)$ for some CNRPQ $\mathbf{q}'$.

Let $\mathbb{D}= (G, \mathcal{D})$ be a heterogeneous database where $G = \{(a, a, a)\}$ where $a$ does not occur in $\mathbf{q}'(?x, ?y)$ and $\mathcal{D} = \{D\}$ with relation $D = \{\langle a, b\rangle\}$. We have $\semm {\mathbf{q}(?x,?y)}{\mathbb{D}} = \emptyset$. By Lemma \ref{lem:ecnrpq-cnrpq}, $\semm {\mathbf{q}'(?x, ?y)}{G}$ is not empty, however, we have arrived a contradiction.

\noindent Proof of \textbf{Lemma} \ref{lem:fcnrpqp-single}\\
By induction on the structure of $\mathbf{q}$.
\begin{compactitem}
\item If $\mathbf{q}(?x, ?y)$ is of the form $R(w_1, \ldots, w_m)$ then $\semm{\mathbf{q}(?x, ?y)}{\mathbb{D}}$ is empty.
\item If $\mathbf{q}(?x, ?y)$ is of the form $(?x, e, ?y)$ where $e$ is an nre-expression then it follows definitions.
\item If $\mathbf{q}(?x, ?y)$ is of the form (\ref{equ:FCNRE}) then, by definition, we conclude that $\semm {\mathbf{q}(?x, ?y)}{G} =  \semm {\varphi}{\mathcal{D}} \Join \semm {(u_1, e_1, v_1)}{G} \Join \ldots \Join \semm {(u_n, e_n, v_n)}{G}$. By the definition of the operator $\Join$ on sets of mappings, $\semm {\mathbf{q}(?x, ?y)}{G}$ contains at most one mapping by induction.
\end{compactitem}
Therefore, $\semm{\mathbf{q}(?x, ?y)}{\mathbb{D}}$ contains at most one mapping.

\noindent Proof of \textbf{Theorem} \ref{thm:uecnrpq-ecnrpq}\\
Consider an FCNRPQ $\mathbf{q}(?x, ?y) = (?x, \nexts, ?y) \vee (?x, \nexts^{-1}, ?y)$. Suppose, for the sake of contradiction, that $\mathbf{q}(?x, ?y)$ is expressible as $\mathbf{q}'(?x, ?y)$ for some FCNRPQ $\mathbf{q}'$.

Let $\mathbb{D}= (G, \emptyset)$ be a heterogeneous database where $G = \{(a, p, b)\}$. We have $\semm {\mathbf{q}(?x,?y)}{\mathbb{D}} = \{(?x \to a, ?y \to b), (?x \to b, ?y \to a)\}$. By Lemma \ref{lem:fcnrpqp-single}, $\semm {\mathbf{q}'(?x, ?y)}{G}$ contains at most one mapping, however, we have arrived a contradiction.

\noindent Proof of \textbf{Proposition} \ref{prop:PP-not-UFCRPQ}\\
Consider a PP query $(?x, !p, ?y)$. Suppose, for the sake of contradiction, that
$(?x, !\iri, ?y)$ is expressible as $\mathbf{q}(?x,?y)$ for some UFCNRPQ query $\mathbf{q}$. Let $\mathbb{D}= (\{(a, q, b)\}, \emptyset)$ and $\mathbb{D}'= (\{(a, q, b)$, $(a, p, b)\}, \emptyset)$. Clearly, $\mathbb{D}\subseteq \mathbb{D}'$. By Lemma \ref{lem:PP-not-UFCRPQ}, $\semm {\mathbf{q}(?x, ?y)}{\mathbb{D}} \subseteq \semm {\mathbf{q}(?x, ?y)}{\mathbb{D}'}$. Since $(?x, !\iri, ?y)$ is expressible as $\mathbf{q}(?x,?y)$ and $\semm {(?x, !p, ?y)}{\{(a, q, b)\}} = \{(?x \to a, ?y \to b)\}$, we can conclude that $(?x \to a, ?y \to b) \in \semm {(?x, !p, ?y)}{\{(a, q, b), (a, p, b)\}}$. $\semm {(?x, !p, ?y)}{\{(a, q, b), (a, p, b)\}} = \emptyset$, however, we have arrived a contradiction.

\noindent Proof of \textbf{Proposition} \ref{prop:RPQ-not-nSPARQL}\\
Consider an RPQ$^{\mathrm{sf}}(\mathrm{\mathsmaller R})$ $\mathbf{q}(?x, ?y) = R(w_1, \ldots, w_m)$. Clearly, there exists no nSPARQL pattern $P$ such that $P$ expresses $\mathbf{q}(?x, ?y)$.

\noindent Proof of \textbf{Proposition} \ref{prop:RPQ-not-nSPARQL-sf}\\
Let $\mathbf{q}(u, v)$ be an RPQ$^{\mathrm{sf}}(|,\mathrm{\mathsmaller N},\wedge,\vee)$. By induction on the structure of $\mathbf{q}$.
\begin{compactitem}
\item If $\mathbf{q}(u, v)$ is an RPQ$^{\mathrm{sf}}(|,\mathrm{\mathsmaller N})$ then there exists always some pattern $P$ in nSPARQL$^{\mathrm{sf}}$ such that $\semm {\mathbf{q}(u,v)}{G} = \semm {P}{G}$ for any RDF graph $G$ (see the proof of \cite[Theorem 4.1]{Taski_SPARQL}).
\item If $\mathbf{q}(u, v)$ is of the form (\ref{equ:CNRE}) then, construct a pattern $P$ as follows: $P = P_1\, \andd\, \ldots \andd\, P_n$ where $P_i$ is an nSPARQL pattern with $\semm {P_i}{G} = \semm {(u_i, e_i, v_i)}{G}$ for any RDF graph $G$ for $i \in \{1,2,\ldots, n\}$. So $\semm {\mathbf{q}(u, v)}{G} = \semm {P}{G}$ by induction.
\item Finally, if $\mathbf{q}(u, v)$ is of the form (\ref{equ:UFCNRE}) then, construct a pattern $P$ as follows: $P = P_1\, \andd\, \ldots \andd\, P_n$ where $P_i$ is an nSPARQL pattern with $\semm {P_i}{G} = \semm {\mathbf{q}(u_i, v_i)}{G}$ for any RDF graph $G$ for $i \in \{1,2,\ldots, n\}$. So $\semm {\mathbf{q}(u, v)}{G} = \semm {P}{G}$ by induction.
\end{compactitem}
Therefore, RPQ$^{\mathrm{sf}}(|,\mathrm{\mathsmaller N},\wedge,\vee)$ is expressible in nSPARQL$^{\mathrm{sf}}$.

\noindent Proof of \textbf{Theorem} \ref{thm:FPQ-SPARQL}\\
By Proposition \ref{prop:RPQ-not-nSPARQL}, RPQ$^{\mathrm{sf}}(\mathrm{\mathsmaller R})$ is not expressible in nSPARQL. Since SPARQL is expressible in nSPARQL, the first holds, that is, RPQ$^{\mathrm{sf}}(\mathrm{\mathsmaller R})$ is not expressible in SPARQL.

Since nSPARQL$^{\mathrm{sf}}$ is expressible in SPARQL (see \cite[Theorem 4.1]{Taski_SPARQL}), the second holds, that is, RPQ$^{\mathrm{sf}}(|,\mathrm{\mathsmaller N},\wedge,\vee)$ is expressible in SPARQL.

\noindent Proof of \textbf{Proposition} \ref{prop:FCNRPQ_complexity}\\
We can reduce the evaluation problem of FCNRPQ in relations to the evaluation problem of the conjunctive first-order queries (CQ) in relations where the data complexity of CQ is in polynomial time and the data complexity of CQ is in NP-complete time \cite{ahv_book}.

\end{document}